\documentclass[a4paper,11pt]{article}
\usepackage{pos}
\usepackage{lineno}
\usepackage{xspace}
\newcommand{\jpsi}{J/$\psi$}
\def\pt{$p_{\mathrm{T}} $\xspace} 

\title{Recent results on heavy flavour in small and large systems from ALICE}
 \ShortTitle{Recent results on HF with ALICE}

\author*[a]{Chun-Lu Huang on behalf of the ALICE Collaboration}

\affiliation[a]{IJCLab, CNRS/IN2P3, Université Paris-Saclay, 91405 Orsay, France}

\emailAdd{chun-lu.huang@cern.ch}

\abstract{Heavy-flavour hadrons, containing open or hidden charm and beauty flavours, are considered as hard probes of the quark--gluon plasma (QGP), which is a hot and dense phase state of quantum chromodynamics (QCD) matter. Production of heavy flavours involves large momentum transfer processes during collisions. Heavy quarks are expected to be produced in the initial stage of collisions and therefore prior to the formation of the QGP in heavy-ion collisions. Consequently, heavy quarks travel through and interact with the QGP. Hence, measuring both open and hidden heavy-flavour production in large systems provides information on the QGP properties and heavy quark hadronisation. Heavy-flavour production is measured in small and large systems with ALICE in order to test (perturbative) QCD and study the effects induced by the medium. In this contribution, recent ALICE results on heavy-flavour production in pp, p--Pb and Pb--Pb collisions are reported.   }

\FullConference{%
  The Eighth Annual Conference on Large Hadron Collider Physics-LHCP2020 \\
  25-30 May, 2020\\
  online}

\begin{document}
\maketitle

In the QCD phase diagram, the quark--gluon plasma (QGP) is a state of matter in which quarks and gluons are deconfined and that requires high temperature and/or high density to be produced. In order to study the properties of the QGP, heavy-ion collisions are used. Heavy-flavour (HF) mesons, such as $D$, $B$ or J/$\psi$, contain at least one charm or beauty quark. The production of heavy flavour in hadron-hadron collisions (small system) is studied with perturbative quantum chromodynamics (pQCD).  In high-energy heavy-ion collisions, where the formation of a QGP could occur, heavy quarks are produced in the initial stage of the collisions and the production of heavy flavour hadrons is affected by medium related effects, which are classified into cold or hot nuclear matter effects. The former effects, for example modification of nuclear parton distribution functions (gluon shadowing at the LHC energies \cite{Kusina:2017gkz}), gluon saturation or parton multiple scattering and energy loss \cite{Arleo_2011}, are related to the modification of heavy-flavour production in p--A (small system) and AA (large system) collisions  in absence of the QGP formation. The hot nuclear matter effects, such as colour screening \cite{Matsui:1986dk}, (re)generation \cite{PhysRevC.63.054905} in the case of hidden HF or energy loss \cite{Nahrgang:2013xaa, Song:2015ykw, He:2014cla, Uphoff:2014hza} in the case of open and hidden HF, can be studied in AA collisions where the formation of a QGP is expected. In this contribution, recent results on heavy-flavour production in small and large systems from A Large Ion Collider Experiment (ALICE) are reported. 


The ALICE detector \cite{Aamodt:2008zz} is able to detect the decay products of heavy flavours in the central barrel or in the muon spectrometer. The Inner Tracking System (ITS), the Time Projection Chamber (TPC) and the Time-Of-Flight system detect and identify hadronic decay products from open heavy flavour hadrons, and electrons from quarkonium decay, in the pseudorapidity range $|\eta|<$ 0.9. The ElectroMagnetic Calorimeter ($|\eta|<$ 0.7) along with the TPC and ITS detectors can be used to measure the production of heavy-flavours via their (semi-) electronic decay. The prompt and non-prompt HF contributions can be separated at mid-rapidity by resolving the secondary vertex from the primary production vertex. At forward rapidity (2.5 $<y<$ 4), the muon spectrometer measures inclusive HF production either in the muon (open HF) or in the dimuon (quarkonia) decay channel.


\begin{figure}[!htbp]
 \begin{center} 
  \includegraphics[scale=0.3]{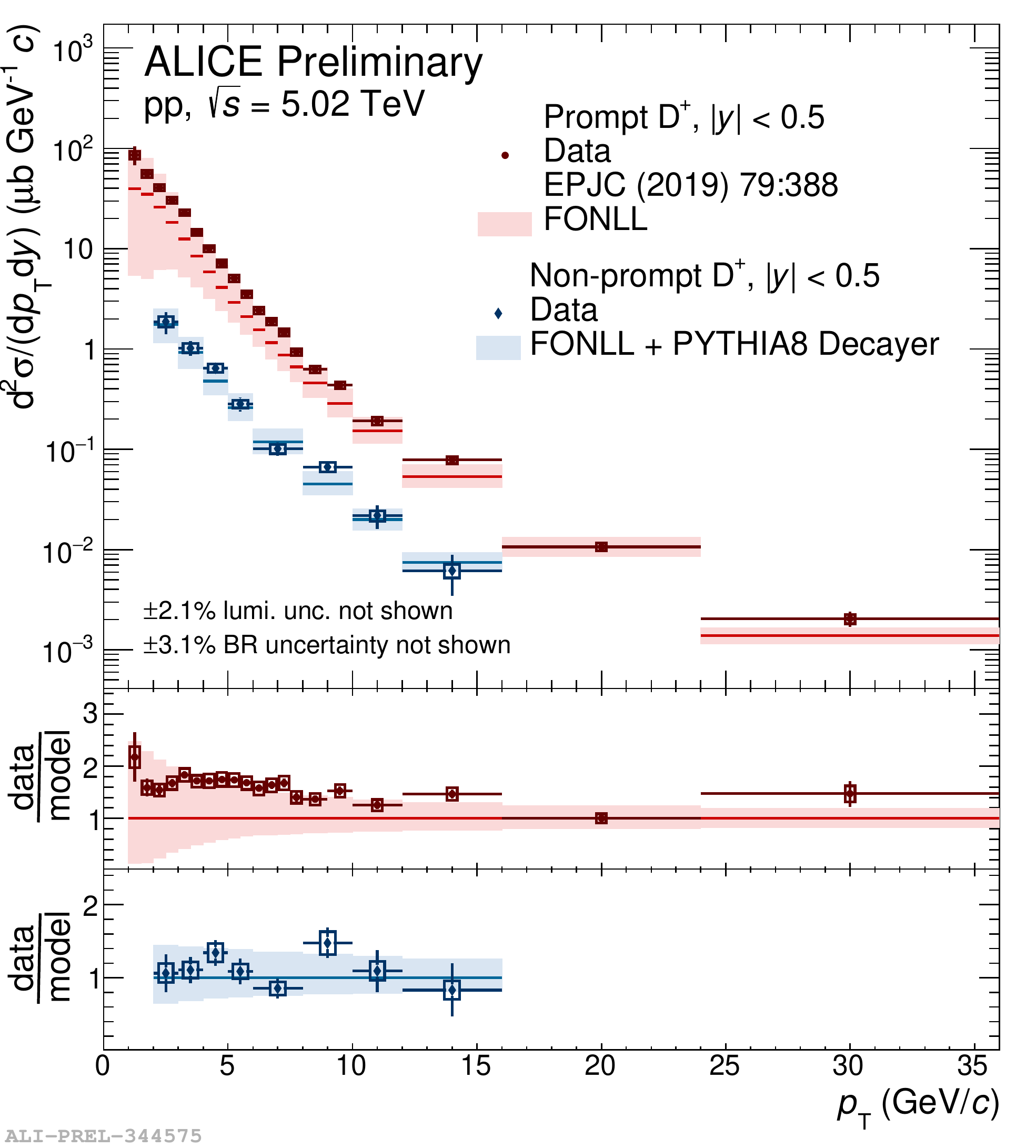}
  \includegraphics[scale=0.44]{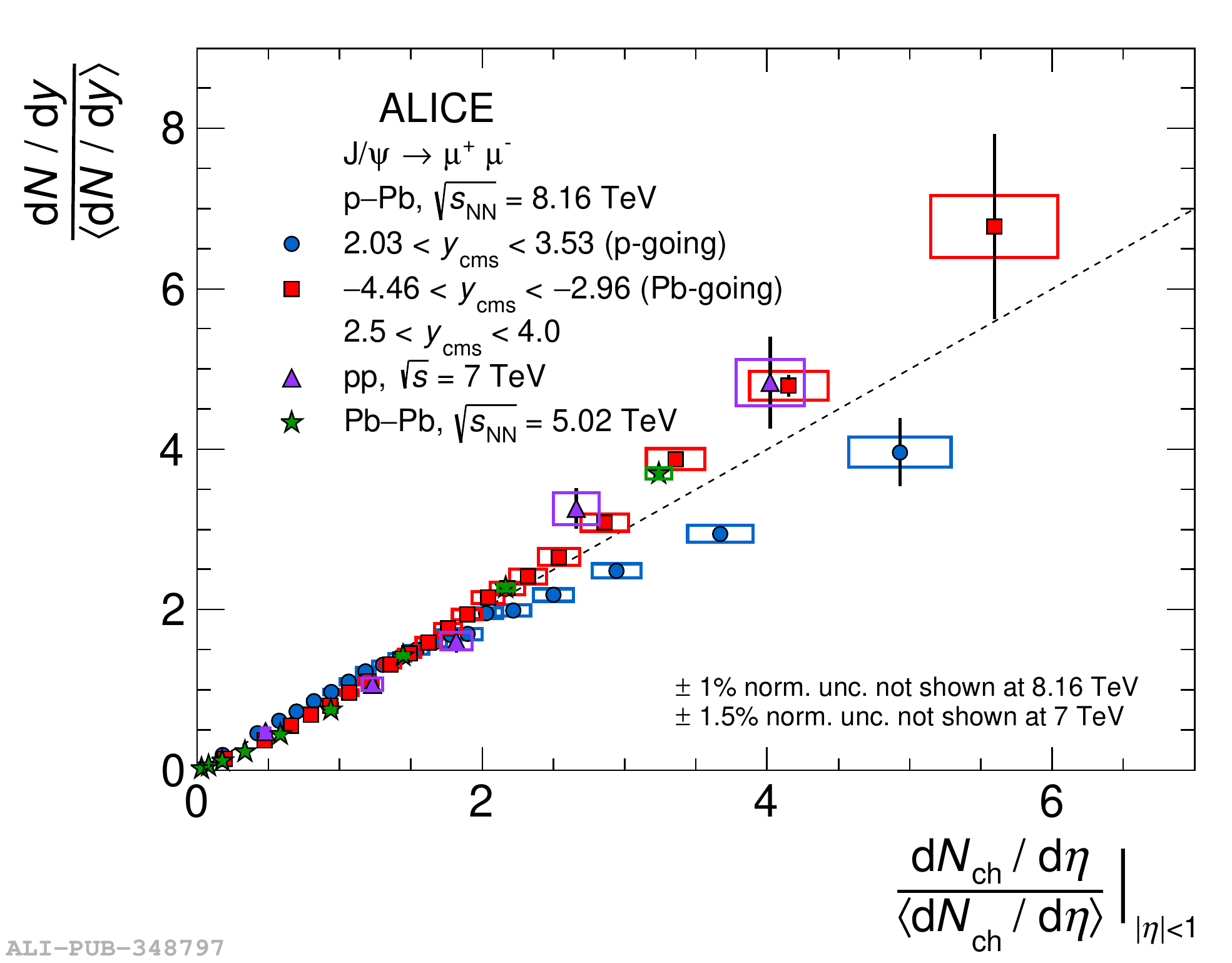}
 \end{center}
 \caption{\scriptsize{Left: $p_{\mathrm{T}}$-differential cross section of prompt \cite{Acharya:2019mgn} and non-prompt $D^{+}$ mesons in pp collisions at $\sqrt{s}$ = 5.02 TeV. Data to theoretical calculation ratios are shown at the bottom of the figure. Right: normalised yield of inclusive \jpsi~as a function of the normalised charged-particle pseudorapidity density in various collision systems at different energies and rapidities \cite{Acharya:2020giw}. See text for details.}}
 \label{fig:open_charm_sigma_pp}
 \end{figure}
Measurements of $D$-meson production in pp collisions are useful to test collinear factorisation within the pQCD framework and provide a reference for the study of the medium effects in heavy-ion collisions. The left panel of Fig.\,\ref{fig:open_charm_sigma_pp} shows the \pt-differential cross section of prompt \cite{Acharya:2019mgn} and non-prompt $D^{+}$, which is from beauty-hadron decay, for $|y|<$ 0.5 in pp collisions at $\sqrt{s}$ = 5.02 TeV. The theoretical calculation based on Fixed-Order Next-to-Leading Logarithm (FONLL) \cite{Cacciari:2012ny} (for the $B$-meson production) and PYTHIA 8 Decayer (for the $B\rightarrow D$ decay kinematics for the non-prompt $D$) is compatible within uncertainties with the non-prompt $D^{+}$ data in the whole measured \pt~range. The prompt $D^{+}$ cross section measured at the same energy is larger than the non-prompt $D^{+}$ one by an order of magnitude for 2 $< p_{\mathrm{T}} <$ 15 GeV/$c$. Besides, the FONLL calculation gives a good description of the prompt $D^{+}$ cross section for 0 $< p_{\mathrm{T}} <$ 35 GeV/$c$.

High-multiplicity events in p--Pb collisions provide an opportunity to probe the collective behaviour in small systems, which might indicate that similar mechanisms are at play in large and small systems. The self-normalised \jpsi~yields are shown as a function of the self-normalised charged-particle multiplicity measured in various collision systems at different energies and rapidities in the right panel of Fig.\,\ref{fig:open_charm_sigma_pp}. In p-Pb collisions at $\sqrt{s_{\mathrm{NN}}}$ = 8.16 TeV \cite{Acharya:2020giw}, the backward rapidity yield grows faster than the forward rapidity one and faster than linear. The self-normalised \jpsi~yield as a function of the multiplicity is also compared to the one in p--Pb collisions at $\sqrt{s_{\mathrm{NN}}}$ = 5.02 TeV \cite{Acharya:2020giw}. The comparison shows good agreement. The mean $p_{\mathrm{T}}$, $\langle p_{\mathrm{T}} \rangle$, is systematically smaller at backward than at forward rapidity \cite{Acharya:2020giw}. The different behaviour as a function of rapidity could be explained by the rapidity dependence of the CNM effects, such as (anti-)shadowing, gluon saturation and coherent energy loss. The comparison between $\sqrt{s_{\mathrm{NN}}}$ = 5.02 TeV and 8 TeV suggests a common origin of the multiplicity trend. Besides, the $\langle p_{\mathrm{T}} \rangle$ saturates at high multiplicity for the two rapidity intervals. At backward rapidity, the simultaneous increase of the yield together with the saturation of $\langle p_{\mathrm{T}} \rangle$ may point to \jpsi~production from an incoherent superposition of parton--parton collisions. In pp collisions at $\sqrt{s}$ = 7 TeV and Pb--Pb collisions at $\sqrt{s_{\mathrm{NN}}}$ = 5.02 TeV, the \jpsi~yield is compatible with the backward rapidity p--Pb results at $\sqrt{s_{\mathrm{NN}}}$ = 8.16 TeV. 
 
 
\begin{figure}[!htbp]
 \begin{center} 
 \includegraphics[scale=0.42]{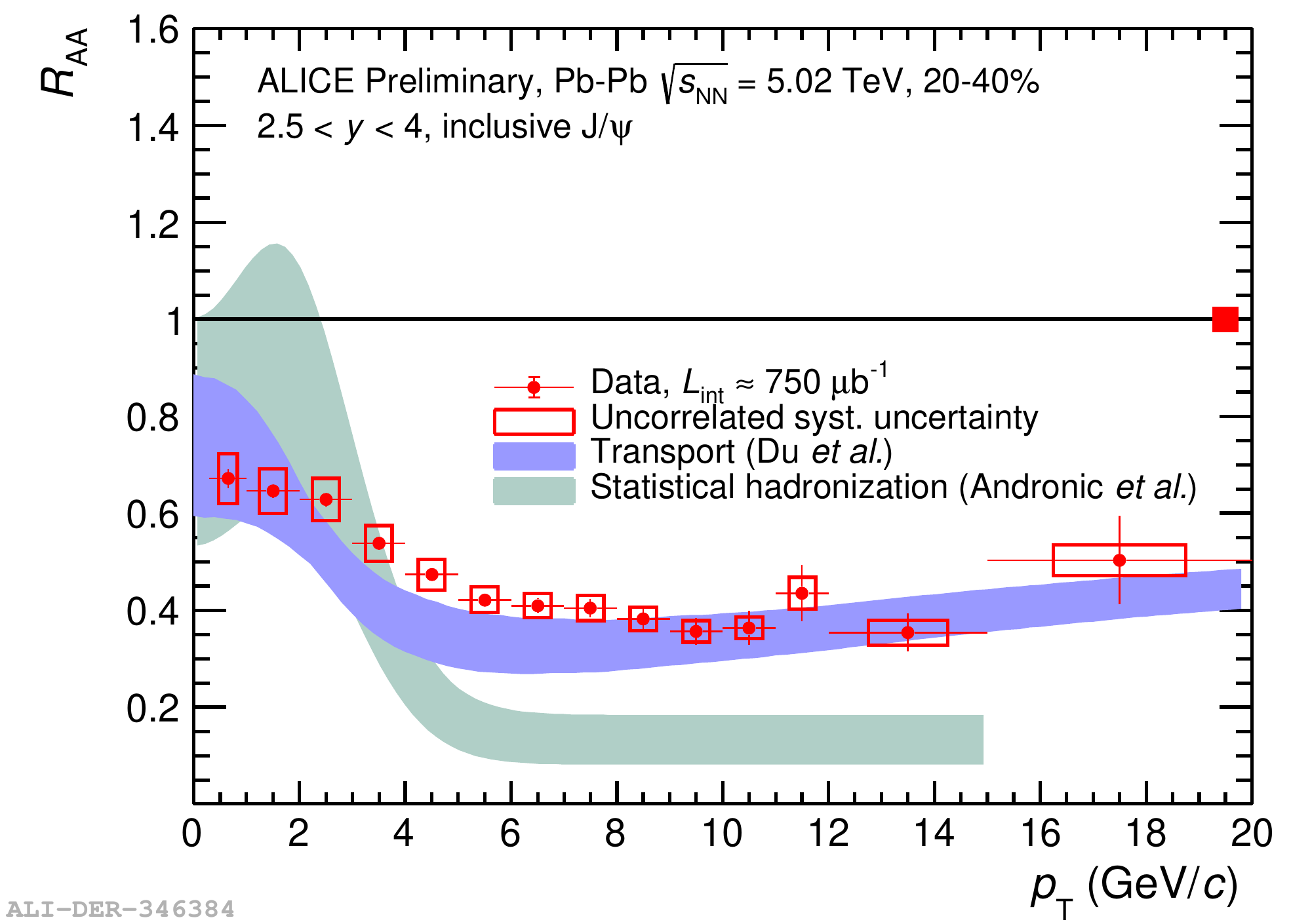}
  \includegraphics[scale=0.09]{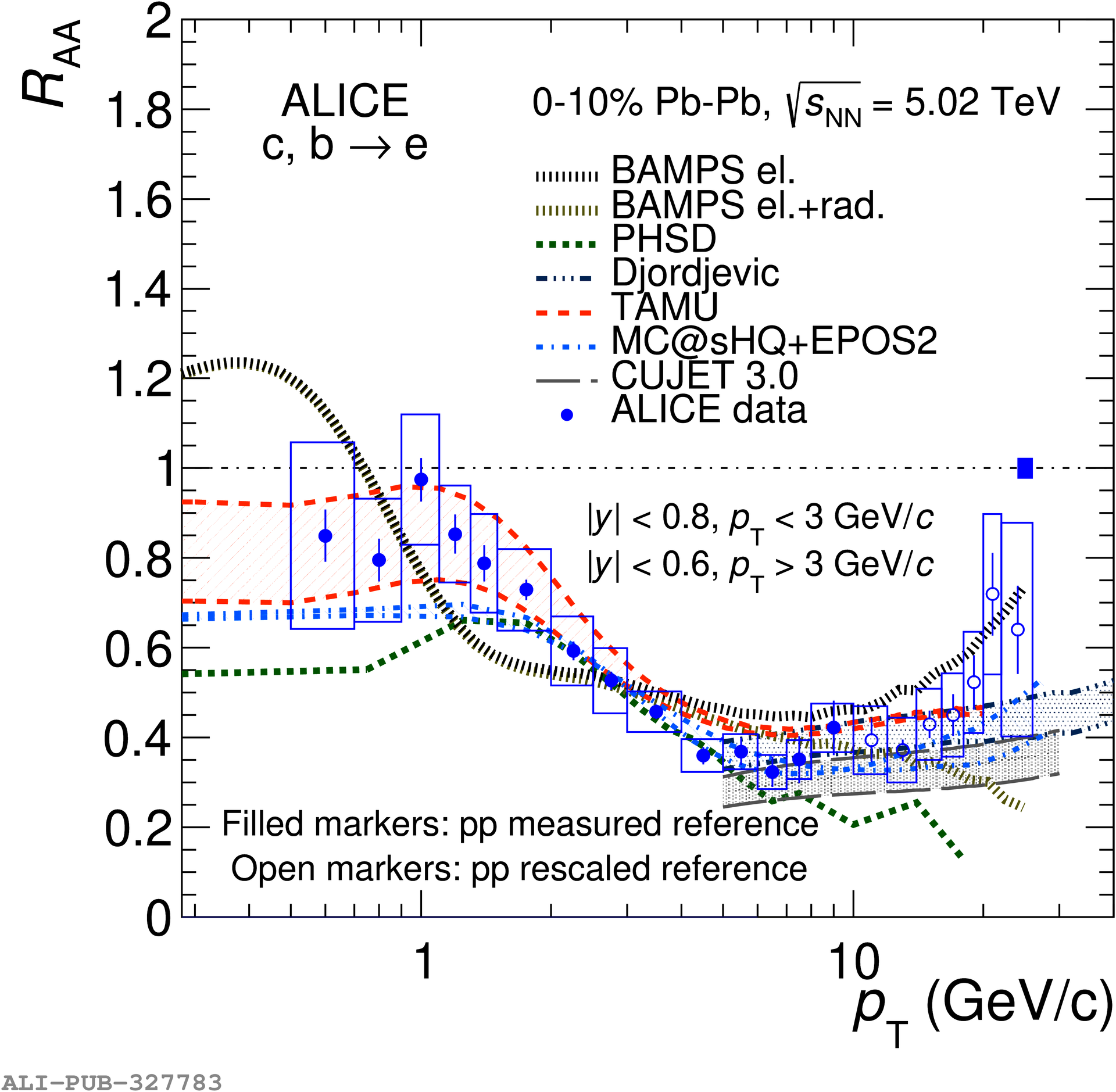}
 \end{center}
 \caption{\scriptsize{Left: inclusive \jpsi~nuclear modification factor as a function of \pt~for 2.5 $ < y < $ 4 in the 20--40\% centrality range in Pb--Pb collisions at $\sqrt{s_{\mathrm{NN}}}$ = 5.02 TeV. Right: nuclear modification factor of electrons from semi-leptonic heavy-flavour hadron decays measured in the 0--10\% centrality range in Pb--Pb collisions at $\sqrt{s_{\mathrm{NN}}}$ = 5.02 TeV \cite{Acharya:2019mom}. See text for details.}}
 \label{fig:jpsi_electron_raa_PbPb}
 \end{figure} 

In AA collisions, heavy quarks interact with the medium and allow us to probe several mechanisms through the measurement of open and hidden HF. The left panel of Fig.\,\ref{fig:jpsi_electron_raa_PbPb} shows the inclusive \jpsi~nuclear modification factor ($R_{\mathrm{AA}}$) as a function of \pt~for 2.5 $ < y < $ 4 in the 20--40\% centrality range in Pb--Pb collisions at $\sqrt{s_{\mathrm{NN}}}$ = 5.02 TeV. The \jpsi~$R_{\mathrm{AA}}$ is suppressed in the whole \pt~range. It increases with decreasing \pt~while it stays flat at around 0.4 for 6 $< p_{\mathrm{T}} < $ 20 GeV/$c$. The data are compared to theoretical calculations based on a transport \cite{Zhao:2011cv} and on the statistical hadronisation model \cite{Andronic:2019wva}. The transport model qualitatively describes the data in the full \pt~range. The transport model includes charmonium (re)generation in the QGP phase for $p_{\mathrm{T}} < $ 6 GeV/$c$ and primordial \jpsi~that survive in the QGP at high $p_{\mathrm{T}}$. The statistical hadronisation calculation is consistent with the data for $p_{\mathrm{T}} <$ 4 GeV/$c$ within the large model uncertainties. In this region, the model considers that the initially produced \jpsi~are fully suppressed and the \jpsi~yield is dominated by the regeneration from the $c\Bar{c}$ pairs at the phase transition. The calculation underestimates the data for $p_{\mathrm{T}} \geq$ 4 GeV/$c$ where \jpsi~are initially produced in the nucleus corona where no QGP effect is expected in the model. The energy loss and colour screening are at play for \pt~$\geq$ 6 GeV/$c$. The right panel of Fig.\,\ref{fig:jpsi_electron_raa_PbPb} shows the \pt-dependence of the $R_{\mathrm{AA}}$ of electrons from semi-leptonic heavy-flavour decays measured in the 0--10\% centrality class in Pb--Pb collisions at $\sqrt{s_{\mathrm{NN}}}$ = 5.02 TeV. The $R_{\mathrm{AA}}$ is suppressed and reaches a minimum at \pt~= 7 GeV/$c$. The $R_{\mathrm{AA}}$ rises up with decreasing $p_{\mathrm{T}}$ starting from \pt~= 7 GeV/$c$. The data are compared to several model calculations \cite{Nahrgang:2013xaa, Song:2015ykw, He:2014cla, Uphoff:2014hza} that consider different assumptions on the mass dependence of energy loss processes, transport dynamics, charm and beauty quark interaction with the QGP, hadronisation mechanisms of heavy quarks in the medium and heavy-quark production cross section in nucleus–nucleus collisions. In this contribution, only few models are described. The MC@sHQ+EPOS2 \cite{Nahrgang:2013xaa}, PHSD \cite{Song:2015ykw} and TAMU \cite{He:2014cla} models take into account the modification of the nuclear parton distribution functions (nPDFs) as well. The BAMPS+rad.\,model \cite{Uphoff:2014hza} incorporates elastic interactions of heavy quarks with the medium and radiative energy loss. The models provide fair descriptions of the data for \pt~$<$ 5 GeV/$c$ apart from BAMPS+rad. The magnitude of the $R_{\mathrm{AA}}$ suppression at low \pt~can be reproduced by the models taking into account the modification of nPDFs. The TAMU models tend to overestimate the data for \pt~$>$ 3 GeV/c, probably due to the missing implementation of the radiative energy loss. On the other hand, the agreement of the TAMU model with data for \pt~$<$ 3 GeV/$c$ confirms the dominance of elastic collisions at low momenta, together with the importance of considering the nPDFs modifications.  

In summary, this contribution reports on recent experimental results on heavy-flavour production measured with ALICE. In pp collisions at $\sqrt{s}$ = 5.02 TeV, the differential cross sections for prompt and non-prompt $D^{+}$ for $|y|<$ 0.5 provide a test of pQCD calculations. The FONLL theoretical calculations give a successful description of the $D$-meson data. In p--Pb collisions at $\sqrt{s_{\mathrm{NN}}}$ = 8.16 TeV, the \jpsi~yields as a function of the charged particle density at mid-rapidity have been measured. The backward-rapidity \jpsi~yields rise faster than the forward-rapidity yields. This behaviour is expected because of the rapidity dependence of the CNM effects. In Pb--Pb collisions at $\sqrt{s_{\mathrm{NN}}}$ = 5.02 TeV, the \jpsi~$R_{\mathrm{AA}}$ as a function of \pt~for 2.5 $<y<$ 4 in the 20--40\% centrality class is measured. The \jpsi~(re)generation contribution dominates for \pt~$<$ 6 GeV/$c$ while there is an interplay between colour screening and energy loss for \pt~$\geq$ 6 GeV/$c$. On the other hand, the $R_{\mathrm{AA}}$ of electrons from semi-leptonic heavy-flavour decay in the 0--10\% centrality class exhibits a suppression that is stronger at high \pt~and well reproduced by models including heavy-quark energy loss in medium. 


\clearpage
\bibliography{my-bib-database.bib}

\providecommand{\href}[2]{#2}\begingroup\raggedright\begin{thebibliography}{10}

\bibitem{Kusina:2017gkz}
A.~Kusina, J.-P.~Lansberg, I.~Schienbein and H.-S.~Shao, \emph{{Gluon Shadowing
  in Heavy-Flavor Production at the LHC}},
  \href{https://doi.org/10.1103/PhysRevLett.121.052004}{\emph{Phys. Rev. Lett.}
  {\bfseries 121} (2018) 052004}
  [\href{https://arxiv.org/abs/1712.07024}{{\ttfamily 1712.07024}}].

\bibitem{Arleo_2011}
F.~Arleo, S.~Peigné and T.~Sami, \emph{Revisiting scaling properties of
  medium-induced gluon radiation},
  \href{https://doi.org/10.1103/physrevd.83.114036}{\emph{Physical Review D}
  {\bfseries 83} (2011) }.

\bibitem{Matsui:1986dk}
T.~Matsui and H.~Satz, \emph{{$J/\psi$ Suppression by Quark-Gluon Plasma
  Formation}}, \href{https://doi.org/10.1016/0370-2693(86)91404-8}{\emph{Phys.
  Lett. B} {\bfseries 178} (1986) 416}.

\bibitem{PhysRevC.63.054905}
R.L.~Thews, M.~Schroedter and J.~Rafelski, \emph{Enhanced \jpsi~production in
  deconfined quark matter},
  \href{https://doi.org/10.1103/PhysRevC.63.054905}{\emph{Phys. Rev. C}
  {\bfseries 63} (2001) 054905}.

\bibitem{Nahrgang:2013xaa}
M.~Nahrgang, J.~Aichelin, P.B.~Gossiaux and K.~Werner, \emph{{Influence of
  hadronic bound states above $T_c$ on heavy-quark observables in Pb + Pb
  collisions at at the CERN Large Hadron Collider}},
  \href{https://doi.org/10.1103/PhysRevC.89.014905}{\emph{Phys. Rev. C}
  {\bfseries 89} (2014) 014905}
  [\href{https://arxiv.org/abs/1305.6544}{{\ttfamily 1305.6544}}].

\bibitem{Song:2015ykw}
T.~Song, H.~Berrehrah, D.~Cabrera, W.~Cassing and E.~Bratkovskaya, \emph{{Charm
  production in Pb + Pb collisions at energies available at the CERN Large
  Hadron Collider}},
  \href{https://doi.org/10.1103/PhysRevC.93.034906}{\emph{Phys. Rev. C}
  {\bfseries 93} (2016) 034906}
  [\href{https://arxiv.org/abs/1512.00891}{{\ttfamily 1512.00891}}].

\bibitem{He:2014cla}
M.~He, R.J.~Fries and R.~Rapp, \emph{{Heavy Flavor at the Large Hadron Collider
  in a Strong Coupling Approach}},
  \href{https://doi.org/10.1016/j.physletb.2014.05.050}{\emph{Phys. Lett. B}
  {\bfseries 735} (2014) 445}
  [\href{https://arxiv.org/abs/1401.3817}{{\ttfamily 1401.3817}}].

\bibitem{Uphoff:2014hza}
J.~Uphoff, O.~Fochler, Z.~Xu and C.~Greiner, \emph{{Elastic and radiative heavy
  quark interactions in ultra-relativistic heavy-ion collisions}},
  \href{https://doi.org/10.1088/0954-3899/42/11/115106}{\emph{J. Phys. G}
  {\bfseries 42} (2015) 115106}
  [\href{https://arxiv.org/abs/1408.2964}{{\ttfamily 1408.2964}}].

\bibitem{Aamodt:2008zz}
{\scshape ALICE} collaboration, \emph{{The ALICE experiment at the CERN LHC}},
  \href{https://doi.org/10.1088/1748-0221/3/08/S08002}{\emph{JINST} {\bfseries
  3} (2008) S08002}.

\bibitem{Acharya:2019mgn}
{\scshape ALICE} collaboration, \emph{{Measurement of ${{\mathrm{D}}^0}$ ,
  ${{\mathrm{D}}^+}$ , ${{\mathrm{D}}^{*+}}$ and
  ${{\mathrm{D}}^+_{\mathrm{s}}}$ production in pp collisions at
  ${\sqrt{{\textit{s}}}~=~5.02~{\text {TeV}}}$ with ALICE}},
  \href{https://doi.org/10.1140/epjc/s10052-019-6873-6}{\emph{Eur. Phys. J. C}
  {\bfseries 79} (2019) 388}
  [\href{https://arxiv.org/abs/1901.07979}{{\ttfamily 1901.07979}}].

\bibitem{Acharya:2020giw}
{\scshape ALICE} collaboration, \emph{{J/$\psi$ production as a function of
  charged-particle multiplicity in p-Pb collisions at $\sqrt{\textit{s}_{\rm
  NN}}~=~8.16$ TeV}},  \href{https://arxiv.org/abs/2004.12673}{{\ttfamily
  2004.12673}}.

\bibitem{Cacciari:2012ny}
M.~Cacciari, S.~Frixione, N.~Houdeau, M.L.~Mangano, P.~Nason and G.~Ridolfi,
  \emph{{Theoretical predictions for charm and bottom production at the LHC}},
  \href{https://doi.org/10.1007/JHEP10(2012)137}{\emph{JHEP} {\bfseries 10}
  (2012) 137} [\href{https://arxiv.org/abs/1205.6344}{{\ttfamily 1205.6344}}].

\bibitem{Acharya:2019mom}
{\scshape ALICE} collaboration, \emph{{Measurement of electrons from
  semileptonic heavy-flavour hadron decays at midrapidity in pp and Pb-Pb
  collisions at $\sqrt{s_{\rm{NN}}}$ = 5.02 TeV}},
  \href{https://doi.org/10.1016/j.physletb.2020.135377}{\emph{Phys. Lett. B}
  {\bfseries 804} (2020) 135377}
  [\href{https://arxiv.org/abs/1910.09110}{{\ttfamily 1910.09110}}].

\bibitem{Zhao:2011cv}
X.~Zhao and R.~Rapp, \emph{{Medium Modifications and Production of Charmonia at
  LHC}}, \href{https://doi.org/10.1016/j.nuclphysa.2011.05.001}{\emph{Nucl.
  Phys. A} {\bfseries 859} (2011) 114}
  [\href{https://arxiv.org/abs/1102.2194}{{\ttfamily 1102.2194}}].

\bibitem{Andronic:2019wva}
A.~Andronic, P.~Braun-Munzinger, M.K.~Köhler, K.~Redlich and J.~Stachel,
  \emph{{Transverse momentum distributions of charmonium states with the
  statistical hadronization model}},
  \href{https://doi.org/10.1016/j.physletb.2019.134836}{\emph{Phys. Lett. B}
  {\bfseries 797} (2019) 134836}
  [\href{https://arxiv.org/abs/1901.09200}{{\ttfamily 1901.09200}}].

\end{thebibliography}\endgroup

\end{document}